\documentclass[showpacs,aps]{revtex4}
\usepackage{graphicx,verbatim}
\begin{document}

\title{Positivity issues for the pinch-technique gluon propagator and their
resolution}
\author{John M. Cornwall\footnote{Email cornwall@physics.ucla.edu}}
\affiliation{Department  of Physics and Astronomy, University of California,
Los Angeles CA 90095}

\begin{abstract}
\pacs{11.15.Tk, 11.15.Kc \hfill UCLA/09/TEP/42}
Although gauge-boson propagators in asymptotically-free gauge theories
satisfy a dispersion relation, they do not satisfy the K\"allen-Lehmann
(K-L) representation because the spectral function changes sign.  We argue
that this is a simple consequence of asymptotic freedom.  On the basis of
the QED-like Ward identities of the pinch technique (PT) we claim that the
product of the coupling $g^2$ and the scalar part $\hat{d}(q^2)$ of the PT
propagator, which is both gauge-invariant and renormalization-group
invariant,  can be factored into the product of the running charge 
$\bar{g}^2(q^2)$ and a term $\hat{H}(q^2)$ both of which satisfy the K-L
representation although their product does not.  We show that this behavior
is consistent with  some simple analytic models that mimic the
gauge-invariant PT Schwinger-Dyson equations (SDE), provided that the
dynamic gauge-boson mass is sufficiently large.  The  PT SDEs do  not
depend directly on the PT propagator through $\hat{d}$ but only through
$\hat{H}$.      
\end{abstract}

\maketitle 

\section{Introduction}

\subsection{Positivity and gluon mass}

A long-standing problem of non-Abelian gauge theories (NAGTs) is the lack of
positivity of the imaginary part of the gauge-boson propagator, violating
the K\"allen-Lehmann (K-L) representation.  This was
first pointed out in an early paper on the  gauge-invariant pinch-technique
(PT) propagator
\cite{corn82}.  Later, many authors found the same behavior in lattice
simulations of the
gauge-dependent and unphysical propagator of   the
Landau gauge \cite{mandog,auog,bowetal} (and references
therein).  Although it is questionable to assign a physical meaning to this
lack of positivity in a gauge-dependent quantity such as the Landau-gauge
propagator, many authors see it as a sign of confinement, since the
propagator of an unconfined field presumably has a normal K-L
representation.  Aubin and Ogilvie \cite{auog} trace it to technical
deficiencies in lattice gauge-fixing procedures.  

In this paper we argue that it is plausible (but unproven) that this lack of
positivity is an elementary consequence of asymptotic freedom, and is
simply resolved by a factorization of the PT propagator into two terms each
of which is both gauge- and renormalization group (RG)-invariant, and each
satisfies the K-L representation.  We construct some simple analytic models
of the PT propagator and vertex that illustrate the necessary positivity
and absence of unphysical singularities, provided that there is a
sufficiently-large dynamical gluon mass.  We also discuss models of the
Analytic Perturbation Theory (APT) type, which can satisfy positivity with
zero gluon mass but still have unphysical behavior.

  Section \ref{notation} briefly covers notation as well as some background
on the PT.
In Section \ref{signprob} we argue that the lack of positivity is a simple
consequence
of asymptotic freedom and the fact \cite{corn82} that the product of the
coupling $g^2$ and the (scalar part of the) gauge-invariant PT propagator
$\hat{d}(q^2)$ is not only gauge-invariant but also  
RG-invariant, independent of the choice of a renormalization point.  This
makes this product a truly physical quantity.  The same is true for the
photon propagator in QED, as has been known for decades, and for the same
reason: The Ward identities of QED or of the PT require that the gluon
vertex function renormalization constant and a wave-function
renormalization constant be the same.

Not every non-perturbative approximation for the propagator can be expected
to satisfy these positivity constraints, in an asymptotically-free gauge
theory.  
In the rest of the paper we construct non-perturbative models that do
satisfy them, provided that there is  a large enough dynamical gluon mass.

 Section \ref{massparam} is an illustration, within the context of an
analytically-soluble model similar to  an earlier \cite{corn82}
one-dressed-loop Schwinger-Dyson equation (SDE) for the
PT propagator, of how the positivity argument above   can only be realized
with a sufficiently large gluon mass.  In the model, provided that the
gluon mass $m$ is
large enough (on the QCD scale $\Lambda$) each factor in the product
$\bar{g}^2\bar{H}$ behaves precisely as would be expected, with no bizarre
behavior coming from non-positivity. But if $m/\Lambda$ is less than a
critical value $m_c/\Lambda$ spurious singularities arise, such as ghost or
spacelike poles in the propagator. We estimate $m_c/\Lambda\simeq 1.2$ in
our study of the one-dressed-loop PT propagator; given the approximations
made there, we believe the range should be from 1 to 1.5 or so.  This
effectively provides, as we will see,  an upper limit to the running charge
at zero momentum:  $\alpha_s(0)\equiv \bar{g}^2(0)/4\pi\leq 0.5-0.7$.  This
is fairly consistent with other determinations from phenomenology
\cite{amn}, studies \cite{corn82,cornhou,papag,abp} of the PT
SDE, and a study of the functional Schr\"odinger
equation \cite{corn136}.

  Section \ref{aptsec} briefly reviews  the evolution of  APT from an
originally
massless form \cite{shirsol} with correct positivity properties, yet showing
unphysical
behavior, to a massive form rather similar to that of \cite{corn82} and the
present paper.  In the original APT positivity was satisfied, although the
PT upper limit on $\alpha_s(0)$ is exceeded,
with $\alpha_s(0)\approx 1/(4\pi b)\approx 1.1$.  This in itself is not
necessarily serious, but what is serious is that the APT running charge,
even though finite at zero momentum, has infinite slope.  A later work
\cite{shirk}  corrects this deficiency by invoking an {\em ad hoc} gluon
mass, in somewhat the same spirit as the PT gluon mass, with results quite
similar to our first model. Other authors
\cite{papnest} have also invoked masses as cutoffs for APT.   

  Section \ref{toyvert}  remarks on the important fact that the vertex SDE
can be
reformulated entirely in terms of the propagator factor $\hat{H}$ with a
positive spectral function and a special half-proper vertex that is both
gauge- and RG-invariant; the original PT propagator $\hat{d}$, with its
positivity violations, never appears.   This reformulation avoids possible
violations of positivity that could allow unphysical vertex behavior.  We
illustrate with an analytic approximation inspired by a one-dressed-loop
toy model \cite{corn99}  of the Schwinger-Dyson equation for the
three-gluon vertex and show that it has a spurious spacelike singularity if
the gluon mass is too small. 

   Section \ref{toynomass} is a discussion of  certain
typical all-order extensions and resummations of the {\em massless} toy
model that still lead to unphysical singularities; these can only be
resolved with a dynamical gluon mass. 

\subsection{The positivity problem and asymptotic freedom}

Write the obvious factorization of the product $g^2\hat{d}$:
\begin{equation}
\label{propform}
g^2\hat{d}(q^2)=\bar{g}^2(q^2)\hat{H}(q^2)
\end{equation}
where $\bar{g}(q^2)$ is the gauge-, scheme-, and
renormalization-point-independent running charge of the PT.     The other
factor $\hat{H}(q^2)$ has the same properties, since the product does.  We
argue that {\em both} factors obey a standard K-L
representation with a positive imaginary part (our metric is such that
$q^2>0$ for timelike vectors):
\begin{eqnarray}
\label{bothkl}
\bar{g}^2(q^2) & = & \frac{1}{\pi}\int_{4m^2}^{\infty}d\sigma \frac{\rho
(\sigma)}{\sigma - q^2-i\epsilon}\\ \nonumber
\hat{H}(q^2) & = & \frac{1}{\pi}\int_{4m^2}^{\infty}d\sigma \frac{\rho_H
(\sigma)}{\sigma - q^2-i\epsilon}
\end{eqnarray}
and $\rho ,\rho_H$ are positive.  The lower limit involves the dynamical
gluon mass $m$, which we discuss later.

We plausibly know the behavior at infinite momentum of both factors in the
product.  Asymptotic freedom tells us that
\begin{equation}
\label{asymptg}
\bar{g}(q^2)\rightarrow_{q^2\to\infty}\frac{1}{b\ln (-q^2/\Lambda^2)}.
\end{equation}
where $b$ is the lowest-order coefficient in the beta-function and $\Lambda$
is the QCD scale.  As for $\hat{H}(q^2)$ there is no reason from
perturbation theory or non-perturbative PT constructions to believe that it
departs from the simple free-field behavior $1/q^2$ for large momentum. 
For example, it is well-known that the PT is equivalent order by order to
the background field Feynman gauge \cite{papbin}, and  old perturbative
calculations in this gauge through two loops \cite{abbott} show that all
large-momentum logarithms are accounted for in the running charge. The
result is that at large momentum the PT propagator vanishes according to
$\hat{d}(q^2)\sim 1/(q^2\ln q^2)$.  But a propagator vanishing more rapidly
than $1/q^2$
implies that the spectral function   in the would-be
K-L representation is necessarily negative somewhere. 
There is, we claim, only an indirect connection---at least for the
gauge-invariant PT propagator---between non-positivity and confinement (a
connection only to the extent that asymptotic freedom implies confinement).
In fact, the gluon is not confined, but screened, in the usual sense that
the string in the adjoint-representation Wilson loop always breaks at
sufficiently large distance even with no adjoint matter fields.  We will 
not attempt any analysis of non-positivity in the Landau-gauge, but it is
likely that even if some non-positivity comes from the Aubin-Ogilvie
\cite{auog} effect, there will still be some residual non-positivity coming
from asymptotic freedom.

What happens at infrared momenta?  It has long been argued
\cite{corn77,corn82} that the QCD gluon should pick up a dynamical mass
that completely preserves local gauge symmetry.  This is consistent with
phenomenology (for example, \cite{amn,bdf}), and      a number of studies of
the PT Schwinger-Dyson equations \cite{papbin2} have found
\cite{corn82,cornhou,papag,abp} a PT pole mass $m$ of order 0.6 GeV.  These
studies also indicate that the mass runs with momentum and should be
denoted $m(q^2)$, consistent with the operator-product expansion result
that  $m^2(q^2)$ vanishes (modulo logarithms) like $\langle
G_{\mu\nu}G^{\mu\nu}\rangle/q^2$ at large momentum \cite{lavelle}.  There
is a large body of lattice-simulation evidence in the Landau gauge
\cite{alex1,alex2,bouc1,bowm1,bouc2,silva1,silva2,bogo1,stern,bogo2,oliv,bowm2,
ilge2,cucch1,cucch2,bouc3,zhang,gong,bims} for a gluon mass of several
hundred MeV.  (The pole mass of the gluon propagator in any gauge is
gauge-invariant and physical, although because it is a timelike pole it is
not easy to determine from lattices simulations.)                   
   We think it plausible, then, that $\hat{H}(q^2)$ has some such form as:
\begin{equation}
\label{fullpropform}
\hat{H}(q^2)=\frac{1}{\hat{m}^2(q^2)-q^2-i\epsilon}
\end{equation}
where $m(q^2)$ is the running mass.  For simplicity and brevity, we use in
this paper a fixed gluon mass; running does not interfere with the main
positivity arguments.  Then $\hat{H}$ is a simple free massive propagator.

\section{A model of the PT inverse propagator}

\subsection{\label{notation} A few words on the pinch technique}

Begin with some notation.
The pinch-technique propagator has the form:
\begin{equation}
\label{scalarprop}
\hat{\Delta}_{\alpha\beta}(q)=P_{\alpha\beta}(q)\hat{d}(q)+\xi
\frac{q_{\alpha}q_{\beta}}{q^4};\;\;P_{\alpha\beta}(q)=-g_{\alpha\beta}+
\frac{q_{\alpha}q_{\beta}}{q^2}.
\end{equation}
The corresponding inverse pinch-technique propagator is:
\begin{equation}
\label{inversept}
\hat{\Delta}^{-1}_{\alpha\beta}(q)=P_{\alpha\beta}(q)[q^2+\hat{\Pi}(q)]
+\frac{1}{\xi}q_{\alpha}q_{\beta}.
\end{equation}
The scalar function $\hat{d}$ is completely independent of the gauge chosen.

The PT is a systematic way of extracting gauge-invariant proper
self-energies, vertices, and the like from gauge-invariant quantities such
as the S-matrix.  The PT propagator, for example, is  not constructed just
from the usual Feynman graphs; it also receives contributions from other
graphs through a so-called ``pinch", in which longitudinal momenta in
numerators, coming from vertices and propagators, trigger Ward identities
leading to the replacement of certain propagators by unity.  This changes
the topology of the graphs where this occurs, and some of these pinch
contributions are readily recognized as contributions of propagator type. 
Although not recognized at the beginning, it is now known
\cite{papbin} that the PT is the same graphical expansion as that of the
background field-Feynman gauge.  Because the whole point of the PT is to
maintain gauge invariance, it is  essential that, when this graphical
expansion is resummed to a dressed-loop or skeleton expansion, all Green's
functions appearing in the skeleton expansion obey the correct Ward
identities.  In the case of the PT these are the naive Ward identities of
QED, with no ghosts.  

It would appear that progress can only be made by solving all possible SDEs
at once, since that will guarantee satisfaction of the Ward identities. 
However, it is possible to find {\em approximate} three- and higher-point
proper vertices that satisfy the Ward identities exactly and that are
expressed solely in terms of the PT propagator itself.  This approximation,
known as the gauge technique, is valid for infrared-dominated phenomena. 
Although we will not give any details here, we have in mind the gauge
technique of Ref. \cite{cornhou}, which gives the following expression
(group
indices suppressed) for the gauge technique proper vertex:
\begin{equation}
\label{cheqn}
\hat{\Gamma}_{\alpha\beta\gamma}(k_1,k_2,k_3)=g_{\alpha\beta}
(k_1-k_2)_{\gamma}-\frac{k_{1\alpha}k_{2\beta}}{2k_1^2k_2^2}(k_1-k_2)^{\mu} 
\hat{\Pi}_{\mu\gamma}(k_3)-[P^{\mu}_{\alpha}(k_1)\hat{\Pi}_{\mu\beta}(k_2)-
\hat{\Pi}_{\alpha}^{\mu}(k_1)P_{\mu\beta}(k_2)]\frac{k_{3\gamma}}{k_3^2}
+cyc.\;perm.
\end{equation}
where the first term on the right is the bare vertex $\hat{\Gamma}^0$.
Here $\hat{\Pi}_{\mu\nu}=P_{\mu\nu}\hat{\Pi}$ is the PT proper self-energy
introduced above. 
This vertex satisfies:
\begin{equation}
\label{ptward}
k_1^{\alpha}\{\hat{\Gamma}_{\alpha\beta\gamma}(k_1,k_2,k_3)-
\hat{\Gamma}^0_{\alpha\beta\gamma}(k_1,k_2,k_3)\}
=\hat{\Pi}_{\beta\gamma}(k_2)-\hat{\Pi}_{\beta\gamma}(k_3)
\end{equation}
no matter what the choice of $\hat{\Pi}$ is.  In consequence, an approximate
but fully gauge-invariant one-dressed-loop SDE for the PT propagator can be
written solely in terms of that propagator.  It is in this sense that we
speak of studying the PT propagator on its own terms, without further
specification of the vertex beyond that of the gauge technique.  In this
paper we use the self-contained  PT propagator equation of Ref.
\cite{cornhou}.  Later we will study a toy model of a ``half-proper"
three-gluon PT vertex SDE in which some factors coming from the propagators
are incorporated in the definition of the vertex, and we arrive at
essentially self-contained vertex SDEs.

\subsection{\label{signprob} Sign problems}

 Define a K-L   function as a real-analytic function with
at most one pole and a cut along the real positive axis, satisfying an
unsubtracted dispersion relation with a positive spectral function.

We conjecture that the PT propagator is the product of two K-L functions. 
The product of  two K-L functions may or may not be a K-L function, but in
the present case we know it cannot be, because the product vanishes faster
than $1/q^2$ near infinity.
  The product of two K-L functions having this property therefore requires a
spectral function that changes sign.
 Multiply together two K-L functions, call them $G_1$ and
$G_2$.  For each function we have:
\begin{equation}
\label{g12}
G_i=\int d\sigma \frac{\rho_i(\sigma )}{q^2-\sigma}\;\;(i=1,2)
\end{equation}
with $\rho_i$ nowhere negative. The product $G_1G_2$ obeys the dispersion
relation:
\begin{equation}
\label{product}
G_1G_2=\int d\sigma \frac{\rho_{1\times 2}(\sigma )}{q^2-\sigma}
\end{equation}
with
\begin{equation}
\label{productrho}
\rho_{1\times 2}(\sigma)=\mathcal{P}\int d\sigma'\frac{\rho_1(\sigma
)\rho_2(\sigma')
+\rho_1(\sigma' )\rho_2(\sigma )}{\sigma -\sigma'}
\end{equation}
which may be negative in places.  

We note parenthetically that it is possible, in certain field theories
involving scalar particles, for the propagator (taken to be K-L)  to have a
single zero between the particle mass and the lowest two-particle
threshhold and for the proper vertex to have a pole at the same place.  In
our case this would correspond to a zero in $\hat{H}(q^2)$ for
$m^2<q^2<4m^2$.  But for propagator models what we use as a criterion for a
critical mass, based on asymptotic freedom, is not to exclude a zero of the
propagator but to exclude an unwanted pole.  It is true that the vertex
models we study define a ``critical" mass by excluding vertex
singularities, but these are not related in any obvious way to zeroes of
$\hat{H}$.  In any case we explicitly exclude the possibility of a zero in
$\hat{H}$ by assumption, and   our techniques show no signs of
such a zero developing.  See \cite{dfh} for further details and references.

\subsection{\label{massparam}  Mass parametrization}

 The conventional approach to the propagator, whether from the PT or
elsewhere, is to calculate in some approximation the proper self-energy,
that is, the inverse propagator.  
The technique of going from the propagator to the inverse propagator (or
{\em vice versa}) reminds us of analytic perturbation theory (APT), which we
discuss in the following section.   APT was used in the NAGT context to
render $\alpha_s(0)$ finite, even in the zero-mass limit.  However,   an
unphysical singularity remains, and in fact mass-improved APT also has a
critical mass $m_c$ that is rather close to the values we give in this
section.  

  We assume that
there is at most one pole in the PT propagator, at $q^2=m^2$, representing
dynamical gluon mass formation, and no   zeroes, and we replace the factor
$\hat{H}$ of Eq.~(\ref{fullpropform}) by the simple massive propagator
$(q^2-m^2+i\epsilon )^{-1}$, where the mass does not run.   Our pretensions
to accuracy in the infrared do not justify saving
the running, so   $m^2$ can be thought of as either the running mass at zero
momentum or the
pole mass, within the accuracy to which we aspire.

Long ago, a form of the pinch technique was used to estimate the dynamical
gluon 
mass \cite{corn82,cornhou}, and interpreted  in the factorized
form outlined above.    In the formulas of \cite{cornhou} we ignore the
running of the mass, and the non-linear integral equation then becomes:
\begin{equation}
\label{oldch}
[g^2\hat{d}(q)]^{-1}=q^2bZ-\frac{ib}{\pi^2}\int d^4k\hat{H}(k)
\hat{H}(k+q)\{q^2+\frac{m^2}{11} \}+C+\dots
\end{equation}
where the constant $C$ summarizes the seagull graph and other
momentum-independent terms, and we also  omit   two-loop contributions. 
Rather than trying to solve this equation we simply replace $\hat{H}$ by a
free massive propagator, yielding:
\begin{equation}
\label{cornhou1}
[g^2\hat{d} (q^2)]^{-1}=q^2bZ+bJ(q^2;\Lambda^2_{UV})\{q^2+\frac{m^2}{11}\}
+C+\dots.
\end{equation}
where $\Lambda_{UV}$ is an ultraviolet cutoff.
Omitted terms are of higher order in a dressed-loop expansion.
The integral $J(q^2;\Lambda_{UV}^2)$ is:
\begin{equation}
\label{cornhou2}
J(q^2;\Lambda_{UV}^2)=\frac{i}{\pi^2}\int d^4k\frac{1}{(k^2-m^2+i\epsilon
)((k-q)^2-m^2
+i\epsilon )};
\end{equation}
and the UV cutoff is defined through the Feynman-parameter representation:
\begin{equation}
\label{jeqn1}
J(q^2;\Lambda_{UV}^2 )= \int_0^1d\alpha \ln \{\frac{m^2-\alpha (1-\alpha
)q^2-i\epsilon}{\Lambda_{UV}^2}\}.
\end{equation} 
By appropriate choice of $Z$ we make the combination $Z+J$ finite,
and define a renormalized integral $J(q^2;\xi )$.  Aside from its
Feynman-parameter form $J$ has a dispersive representation:
\begin{equation}
\label{jeqn}
J(q^2;\xi )= \int_0^1d\alpha \ln \{\frac{m^2-\alpha (1-\alpha )q^2-i\epsilon
}
{\xi}\}=-q^2\int_{4m^2}^{\infty}\frac{d\sigma}{\sigma}\sqrt{1-\frac{4m^2}
{\sigma}}\frac{1}{\sigma -q^2 -i\epsilon}+\ln (\frac{m^2}{\xi}).
\end{equation}
The PT inverse propagator in finite terms is:
\begin{equation}
\label{finiteprop}
[g^2\hat{d} (q^2)]^{-1}=q^2b\tilde{Z}+b(q^2+\frac{m^2}{11})J(q^2;\xi )+C
+\dots
\end{equation}
Note that this is consistent with the renormalization-invariance of
Eq.~(\ref{propform}).
We make a choice of $\xi$ (or $\tilde{Z}$) that defines what we mean by
$\Lambda$, the finite QCD scale, by requiring that 
\begin{equation}
\label{lambda}
\hat{d}^{-1}(q^2)\rightarrow bg^2q^2\ln (\frac{-q^2}{\Lambda^2})[1+o(1)]
\end{equation}
as $q^2$ approaches infinity in any direction.  (The non-leading
terms are $\mathcal{O}(\ln \ln q^2)$ and can affect the definition of
$\Lambda$ at any particular momentum, but since we deal here only with
one-dressed-loop quantities we cannot use such higher-order terms in the
analysis; one should think of $\Lambda$ as applying to a specific range of
large but finite momenta and that effectively incorporates terms not
vanishing at infinity.)  We choose:
\begin{equation}
\label{xichoice}
\xi = e^{-2}\Lambda^2
\end{equation}
and then we can set $\tilde{Z}=0$.

 Since the inverse propagator is to vanish at $q^2=m^2$ we can
eliminate $C$ by writing:
\begin{equation}
\label{cornhou3}
[g^2\hat{d}(q^2)]^{-1}= b\{J(q^2;\xi )(q^2+\frac{m^2}{11})
-J(m^2; \xi )\frac{12m^2}{11}\}.
\end{equation}
This can be written in dispersive form by using Eq.~(\ref{jeqn}):
\begin{equation}
\label{cornhou4}
[g^2\hat{d}(q^2)]^{-1}=b(q^2-m^2)\{2+\ln
(\frac{m^2}{\Lambda^2})-\int_{4m^2}^{\infty}
d\sigma \sqrt{\frac{\sigma -4m^2}{\sigma}}
 \frac{1}{\sigma -q^2-i\epsilon}[\frac{q^2}{\sigma}+\frac{12m^2}{11(\sigma
-m^2)}]\}.
\end{equation}

We now assume, as discussed in connection with
Eq.~(\ref{fullpropform}), that
\begin{equation}
\label{propcharge}
g^2\hat{d}(q^2)=\frac{\bar{g}^2(q^2)}{q^2-m^2+i\epsilon}
\end{equation}
which yields:
\begin{equation}
\label{propcharge2}
[b\bar{g}^2(q^2)]^{-1}=2+\ln (\frac{m^2}{\Lambda^2})-\int_{4m^2}^{\infty}
d\sigma \sqrt{\frac{\sigma -4m^2}{\sigma}}
 \frac{1}{\sigma -q^2-i\epsilon}[\frac{q^2}{\sigma}+\frac{12m^2}{11(\sigma
-m^2)}].
\end{equation} 

Provided that $m/\Lambda$ is sufficiently large, a condition that we will
investigate below and assume for now, the dispersion relation for the
propagator has the properties discussed in Sec.~\ref{signprob} above.  We
write:
\begin{equation}
\label{g2disp}
\bar{g}^2(q^2)=\frac{1}{\pi}\int_{4m^2}^{\infty}d\sigma \frac{\rho (\sigma
)}{\sigma - q^2-i\epsilon}
\end{equation}
with $\rho$ a  function easily read off from Eq.~(\ref{propcharge2}); we
need not record it explicitly.
This equation shows that  $\rho (\sigma )$ is positive, as we expect. 
 An elementary calculation shows that the dispersion relation for
$-g^2\hat{d}(q^2)$, as taken from Eq.~(\ref{propcharge}), is:
\begin{equation}
\label{newdisp}
-g^2\hat{d}(q^2)= \frac{R}{m^2-q^2-i\epsilon} -\frac{1}{\pi}
\int d\sigma \frac{\rho (\sigma )}{(\sigma -m^2)(\sigma -q^2-i\epsilon)}
\end{equation}
where $R$ is a positive residue, the on-shell value of $\bar{g}^2$:
\begin{equation}
\label{defr}
R=\frac{1}{\pi}\int d\sigma \frac{\rho (\sigma )}{\sigma -m^2}.
\end{equation}
Although each of the two factors $\bar{g}^2(q^2)$ and $1/(m^2-q^2)$ have
positive imaginary parts, the imaginary part of their product, which is:
\begin{equation}
\label{improd}
Im[-g^2\hat{d}(\sigma)]=R\delta (\sigma -m^2)-\frac{\rho (\sigma )}{\sigma
-m^2}
\end{equation}
has one positive term from the pole and another term from the cut that is
everywhere negative.  This is, of course, required by the  large-$q$
behavior of the product which requires that the integral of the imaginary
part vanish.

One implication of the dispersive form in Eq.~(\ref{propcharge2})  is that
$\bar{g}^2$ is positive everywhere where it
is real, that is, in the region $-\infty < q^2 < 4m^2$.
 Another is that
$\bar{g}^2$ is monotonically increasing as $q^2$ decreases.  Since we
expect $\bar{g}^2$ to be monotonically decreasing as the mass increases,
there is a critical mass $m_c$ such that if the physical mass exceeds $m_c$
there is no spurious pole, while there is such a pole if $m<m_c$.  This
critical mass is determined by positivity of the running coupling just
below the threshhold and yields $m_c/\Lambda = \exp [\sqrt{3}\pi/33]
=1.18$.

All the integrals in Eq.~(\ref{propcharge2}) can be evaluated, giving the
running charge explicitly.  In the  regime  $0<q^2 <4m^2$ the explicit
 result is:
\begin{eqnarray}
\label{runchargej}
[b\bar{g}^2(q^2)]^{-1} & = & \frac{1}{q^2-m^2+i\epsilon}
\{ [q^2+\frac{m^2}{11}]J(q^2;\xi )-\frac{m^2}{12}J(m^2;\xi )\} \\ \nonumber
~ & = &\ln(\frac{m^2}{\Lambda^2})   +
        \frac{2}{q^2 - m^2}\{(q^2 + \frac{m^2}{11})\sqrt{\frac{4m^2 -
q^2}{q^2}}\arctan \sqrt{\frac{q^2}{4m^2 - q^2}}- \frac{2\sqrt{3}\pi}{11} \}
\end{eqnarray}
which has an obvious analytic continuation to other regimes.

\begin{figure}
\includegraphics[width=5in]{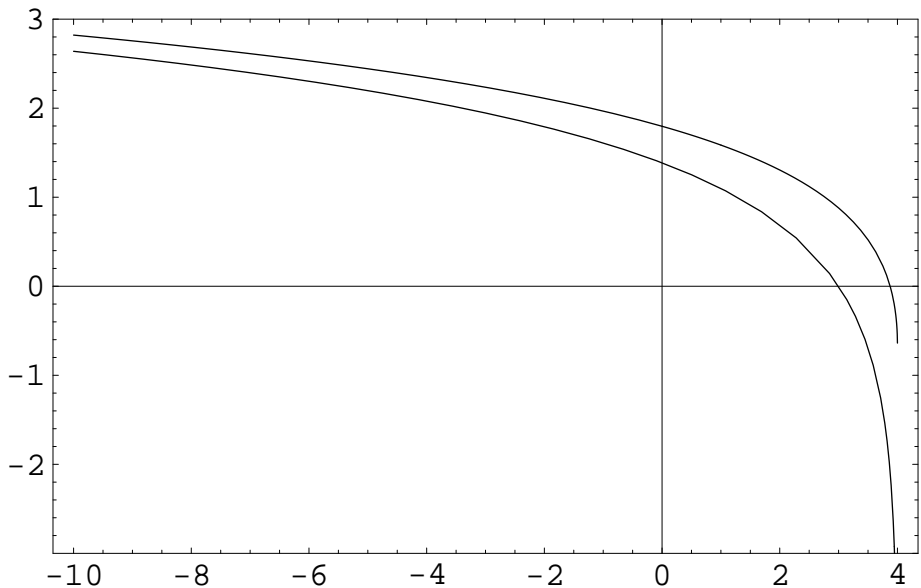}
\caption{\label{fig1} Comparison of new and old expressions for
$[b\bar{g}^2(q^2)]^{-1}$ below threshhold.  The upper curve is the new
[Eq.~(\ref{propcharge2})] and the lower the old [Eq.~(\ref{bestest})]
expression at $m=\Lambda$; the $x$-axis is $q^2/m^2$.  }
\end{figure}

Let us compare this result for the running charge to the 
  old {\em ansatz} of \cite{corn82}, which is:
\begin{equation}
\label{bestest}
 [b\bar{g}(q^2)]^{-1} =   
\ln [(4m^2-q^2-i\epsilon )/\Lambda^2].
\end{equation}
  It is clear that this is not
accurate for the above-threshhold
region $q^2\geq 4m^2$, because it has a pole at $q^2=4m^2-\Lambda^2$, which
is timelike,   provided that $m>\Lambda /2$, and lies below threshhold in
the region where the running charge is real.  However, this expression does
not differ very much from the improved PT
value of Eq.~(\ref{runchargej}) above in the spacelike regime.  In
Fig.~\ref{fig1} we compare the
old expression of Eq.~(\ref{bestest}) 
and the new expression in Eq.~(\ref{propcharge2}) plotted {\em vs.}
$q^2/m^2$, at a
mass ratio $m/\Lambda = 1$.  For any other value, simply add $\ln
(m^2/\Lambda^2)$ to both expressions.  They differ by about 15\% or less
from their average in the
spacelike regime but separate increasingly for $0<q^2<4m^2$ as the 1982
expression approaches its timelike divergence.  

For most phenomenological applications one is interested in the running
charge at a small spacelike momentum transfer.  Either the new expression
from Eq.~(\ref{propcharge2}) or the old one from Eq.~(\ref{bestest}) shows
that $\alpha_s(0)$ increases as $m$ decreases.  A single formula applies to
both cases, with one parameter $\rho$ whose value depends on whether the
new or old expression is used.  We find:
\begin{equation}
\label{ptalpha}
\alpha_s(0)=\{\frac{1}{4\pi b}\}\frac{1}{\ln (\frac{\rho^2
m^2}{\Lambda^2})}
\end{equation}
which is positive provided that $m/\Lambda \geq \rho^{-1}$.
For the new expression:
\begin{equation}
\rho_{new}^{-1}= \exp [-1+\frac{12}{11}(1-\frac{\sqrt{3}\pi}{6} )]\approx
0.41
\end{equation}
and for the old fit $\rho_{old}^{-1} = 0.5$---not much different.  If
$m/\Lambda\geq\rho^{-1}$ the squared running charge is positive for all
spacelike ($q^2<0$) momenta.
 
  What happens when $m<m_c$?  Is this unacceptable, or is there a fix
through
APT?

\section{\label{aptsec} APT and other models}

Massless APT \cite{shirsol}  begins with ordinary perturbation theory
for the running charge, which at one loop is:
\begin{equation}
\label{1loop}
F(q^2)\equiv [\bar{g}(q^2)_1]^{-1} = b\ln
(\frac{-q^2-i\epsilon}{\Lambda^2}).
\end{equation}
The tachyonic pole at $q^2=-\Lambda^2$ is removed using a
renormalization-group-improved extension of fifty-year-old techniques
that impose correct analyticity properties on certain gauge-invariant
quantities such as the Adler D-function or the photon
propagator, which   amounts to postulating the dispersion relation:
\begin{equation}
\label{mlessapt}
\bar{g}^2(q^2)_{APT}=\frac{1}{\pi}\int_0^{\infty}d\sigma 
\frac{-Im\;F}{|F|^2}\frac{1}{\sigma -q^2-i\epsilon}. 
\end{equation}
 As needed, the imaginary part of $\bar{g}^2_{APT}$ is positive.  Since the
dispersion integral can have (by fiat) no pole, the simple result is:  
\begin{equation}
\label{apt}
\alpha_{s,APT}(q^2)= \frac{1}{4\pi b}\{\frac{1}{\ln [(-q^2-i\epsilon )
/\Lambda^2]}+\frac{\Lambda^2}{\Lambda^2+q^2}\}
\end{equation}
with a zero-momentum value of $\alpha_s(0)=1/(4\pi b)\simeq 1.4$ (for
three-flavor QCD). 
Higher-order
renormalization-group improvement changes this value only slightly.  
While this value for $\alpha_s(0)$ is certainly in the right ballpark, there
is an uncomfortable flaw in APT.  It predicts that the slope
$d\alpha_s(q^2)/dq^2$ at $q^2=0$ is negative infinity, which is certainly
unphysical.
Consequently the predicted value in Eq.~(\ref{apt}) is not reliable,
although that is not our main concern here. 

Obviously this massless APT treatment can be trivially extended to the old
proposal  of Eq.~(\ref{bestest}), but this just transfers the infinite
slope to threshhold.  This is a fault to be associated with the proposed
running charge, which is simply not physical near threshhold.

Shirkov later \cite{shirk} proposed to put in, by hand, a gluon mass in a
different way that actually is close in spirit to the version of the PT we
use here. 
Although Shirkov's work is unclear on some relatively insignificant
details, when quarks are omitted it is essentially equivalent to the
following expression: 
\begin{equation}
\label{shireq}  
[b\bar{g}^2(q^2)]^{-1}=J(q^2;\xi ).
\end{equation}
Just as in perturbation theory, $J(q^2;\xi )$ may have an unphysical pole
coming from a zero of $J$; in the massive case with $m/\Lambda >2$ this
pole lies in the region $\Lambda^2<q^2<4m^2$.  The locus of zeroes in
$J(q^2;\xi )$   in Eq.~(\ref{jeqn}) is:
\begin{equation}
\label{zeroloc}
\ln (\frac{4m^2}{\xi})=2\{1-\gamma^{-1} \arctan \gamma\}
\end{equation}
where
\begin{equation}
\label{gamma}
\gamma =\sqrt{\frac{q^2}{4m^2-q^2}}.
\end{equation}
This yields   $m_c\Lambda=1$.  
Of course, it might be possible to remove this singularity for $m<m_c$ by
the same
techniques used for massless APT, with the dispersion relation:
\begin{equation}
\label{aptdisp}
b\bar{g}^2(q^2)=\frac{-1}{\pi}\int_{4m^2}^{\infty}d\sigma \frac{Im\;J(\sigma
;\xi )}{|J(\sigma ;\xi )|^2(\sigma -q^2 -i\epsilon)}.
\end{equation}
However, from the expression for $J$ above threshhold:
\begin{equation}
\label{jabove}
J(q^2;\xi )=\ln (\frac{m^2}{\Lambda^2})+\gamma^{-1}\ln\{\frac{\gamma
+1}{\gamma -1}\}
-i\pi \gamma^{-1}
\end{equation}
one sees that at the critical mass both the real and the imaginary parts of
$J$
vanish at threshhold, leading to a singular running charge at threshhold. 
Presumably this is unphysical.  There are no singularities for larger
values of $m/\Lambda$, so it appears that for mass-improved APT there is a
critical mass:  $m_c/\Lambda = 1$.

\section{\label{toyvert} Toy vertex models and positivity}

It is much too difficult to consider the full
Schwinger-Dyson equations even at the lowest loop level for NAGTs, so we
construct an analytically-soluble toy model.  This new model is in the
spirit of the old toy model of Ref.~\cite{corn99},  which is not
analytically-soluble.    Both models  have the same  large-momentum
behavior, showing    asymptotic freedom and a beta-function with all terms
negative and with factorial growth.

Both models exploit the fact that,  just as the product
$g^2\hat{d}(q^2)$ is not only gauge-invariant but renormalization-group
invariant, there is a similar combination for the PT vertex.
Introduce the notation:
\begin{equation}
\label{znot}
\hat{d}(q^2)=\hat{H}(q^2)\hat{Z}^{-1}(q^2).
\end{equation} 
With our factorization conjecture for the PT propagator this is equivalent
to:
\begin{equation}
\label{newnot}
\bar{g}^2(q^2)=\frac{g^2}{\hat{Z}(q^2)}.
\end{equation}
Because $\hat{g}^2(q^2)$ is positive for spacelike (negative) $q^2$, so is
$\hat{Z}(q^2)$.
Call the proper vertex function in
the pinch technique  $\hat{\Gamma}_{\mu\nu\alpha}^{abc}
(q_1,q_2,q_3)$.  This vertex function, which like the PT propagator is
gauge-invariant and process-independent, obeys a  Ward identity of QED
type, with no contributions from ghosts, schematically of the form
$q_1\cdot \Gamma =\hat{Z}(q_2)-\hat{Z}(q_3)$.   The gauge-invariant and
renormalization-group invariant we call $\hat{G}$ (irrelevant group and
spin indices
omitted):
\begin{equation}
\label{vertform}
\hat{G}(q_1,q_2,q_3)=\frac{g\hat{\Gamma}(q_1,q_2,q_3)}{(\hat{Z}(q_1)\hat{Z}(q_2)
\hat{Z}(q_3))^{1/2}}.
\end{equation}
where $\hat{\Gamma}$ is the PT proper vertex function (again, irrelevant
indices omitted) and $\hat{Z}$ is a factor
in the propagator, as given in Eqs.~(\ref{znot},\ref{newnot}). When all the
momenta are
$\mathcal{O}(q)$ the Ward identity tells us that $\hat{\Gamma}\sim
g^2\bar{g}^{-2}$ at large momentum, and then Eq.~(\ref{vertform}) shows that
$\hat{G}\sim \bar{g}(q)$, as would be expected for a gauge-invariant vertex
function.

One might think that the PT  Schwinger-Dyson equation (SDE)
for $\hat{\Gamma}$ explicitly involves the PT propagator $\hat{d}$
that has a non-positive imaginary part.  Instead we remark   that this
equation can be rewritten in terms only
of the normal propagators $\hat{H}$ and the special vertex $\hat{G}$.  This
is important because $\hat{d}(q^2)$ itself violates the K-L
representation, and if  the skeleton graphs of the SDE were to be modeled
by replacing bare propagators by $\hat{d}$  there  could possibly be
positivity problems in the SDE arising from the $\hat{d}$ terms.  
Schematically the one-dressed-loop SDE is:
\begin{equation}
\label{vertsde}
\hat{G}=\hat{G}_0+\int \hat{G}^3\hat{H}^3+\dots
\end{equation}
Here $\hat{G}$ is the Born term, behaving like $(\ln q^2)^{-3/2}$ when all
momenta are large and $\mathcal{O}(q)$. Note that this equation is {\em
independent} of
the coupling constant $g$, as it must be if $\hat{G}$ is
renormalization-invariant; this independence holds for all vertex skeleton
graphs.  We can now draw conclusions based on the (Euclidean) positivity of
the $\hat{H}$ propagators without fear of difficulties arising from
non-positivity of the spectral function for the propagator itself.  

So far we have not considered numerator factors.  Neither toy model
has them, but each roughly accounts for them by
dropping one of the propagator factors in Eq.~(\ref{vertsde}).
 We know that $\hat{G}\sim (\ln q^2)^{-1/2}$ at large $q$, which dominates
over the inhomogeneous Born term.  Consequently, in the toy model this
inhomogeneous term is dropped.
  Furthermore   in the toy model, $\hat{G}$ depends on only one
momentum, and
only the one-loop skeleton graph is saved.  The original toy model  equation
\cite{corn99} is then:
\begin{equation}
\label{toyd4}
\hat{G}(q)=\frac{ib}{2\pi^2}\int
\frac{d^4k}{(k^2-m^2+i\epsilon )[(q+k)^2-m^2+i\epsilon ]}\hat{G}^3(k) 
\end{equation}
where $b$ is the usual (no-quark) one-loop coefficient in the beta-function.   
This is a universal equation for any coefficient $b$ in the
beta-function, as one sees by using the vertex $R\equiv b^{1/2}\hat{G}$ in
place of $\hat{G}$.  

We will work in Euclidean space, defining $Q^2\equiv -q^2$ as the Euclidean
square of the momentum, positive for spacelike $q$.  Then the kernel of the
non-linear integral equation (\ref{toyd4}) is positive.  If $m\neq 0$ the
kernel is nowhere singular, but if $m=0$ the kernel is singular at zero
momentum.  One implication is that the {\em massless} $\hat{G}$ is
necessarily 
zero or singular   at zero momentum, already suggesting the necessity of a
mass.  We briefly review these facts for the massless model.

\subsection{Massless toy model}

The massless toy model can be converted \cite{corn99} to a differential
equation:
\begin{equation}
\label{massless}
\hat{G}_{tt}+\hat{G}_t=-\frac{b}{2}\hat{G}^3
\end{equation}
where the subscripts indicate derivatives with respect to the variable
$t\equiv \ln (Q^2/\Lambda^2)$  and $\Lambda$ is the usual QCD
mass scale.   We would like to impose physically-sensible boundary
conditions at $Q^2=0$, or $t=-\infty$, but   this is impossible: The
massless version of Eq.~(\ref{toyd4})
is singular at zero momentum unless $\hat{G}(0)=0$, which we forbid. 
Nevertheless the
differential equation
can be solved, showing features expected from perturbation theory, and we
can impose boundary conditions at $t> 0$ and study the ultraviolet behavior.

In the ultraviolet regimes one finds results
familiar from   the RG: The functional form of the
asymptotic vertex
is precisely that of the full NAGT except for the value of some numerical
coefficients, and all signs
agree with what is needed for asymptotic freedom.   
 For large $t$  the second derivative term is non-leading; if dropped, the
general
solution to the first-order differential equation is:
\begin{equation}
\label{gensol}
\frac{1}{\hat{G}^2(t)}-bt=const.
\end{equation}
   This coupling is singular, as massless
perturbation theory must be.  
When the second derivative term is kept, a solution is generated which has
all the same terms as the all-order perturbative running charge in true QCD,
but with somewhat different coefficients.  All these coefficients have the
correct sign for an asymptotically-free theory.  

There is also an interesting beta-function, governed by its own differential
equation.  This comes from the relation:
\begin{equation}
\label{beta}
\hat{G}_t=\frac{1}{2}\beta (\hat{G})
\end{equation}
plus the equation (\ref{massless}) for the vertex, and is:
\begin{equation}
\label{nomassbeta}
\beta(g)[1+\frac{1}{2}\frac{d\beta}{dg}]=-bg^3.
\end{equation}
It was shown \cite{corn99} that   this beta-function
behaves like $-g\sum N!(bcg^2)^N$ for some positive constant $c$,
qualitatively just the same as in any asymptotically-free NAGT, and that the
beta-function solving Eq.~(\ref{nomassbeta}) is singular at a finite upper
critical coupling $g_c$. 

\subsection{Massive toy model}

We give here a new   toy model inspired by but differing slightly from the
original massive toy model of Eq.~(\ref{toyd4}).  The new model is probably
about as accurate as the original in modeling the true SDE, but it can be
analytically solved and shows a critical mass value.  It has the same
ultraviolet (massless) properties as the old model and as QCD does itself.
As before we take $\hat{H}$ as a
 free massive propagator.     There is no longer a simple differential
form of the original toy model equation Eq.~(\ref{toyd4}) when masses are
included, but the following approximation to the Euclidean angular
integration does give an ordinary differential equation:
\begin{equation}
\label{angapprox}
\int \frac{d\Omega_K}{2\pi^2} \frac{1}{[(Q+K)^2+m^2]}\approx
\frac{\theta
(Q^2-K^2)}{Q^2+m^2}
+\frac{\theta (K^2-Q^2)}{K^2+m^2}
\end{equation}
The approximation is exact for large $Q,K$ as well as when $Q>0,K=0$ (or
$K>0,Q=0$) and
is otherwise too large by a factor which is at most about 1.3 times   the
true angular integral at $Q=K=m$.  From this follows
the one-dimensional integral equation:
\begin{eqnarray}
\label{approxint}
\hat{G}(Q^2) & = & J_1(Q^2)+J_2(Q^2)\\ \nonumber
J_1(Q^2) & = &\frac{b}{2(Q^2+m^2)}\int_0^{Q^2}dK^2  
\frac{K^2\hat{G}^3(K^2)}{K^2+m^2} \\ \nonumber
J_2(Q^2) & = & \frac{b}{2}\int_{Q^2}^{\infty}dK^2
\frac{K^2\hat{G}^3(K^2)}{(K^2+m^2)^2} \\
\end{eqnarray}
Every solution of this integral equation $\hat{G}$ satisfies 
the  differential
equation:
\begin{equation}
\label{peqn}
[(Q^2+m^2)\hat{G}(Q^2)]''=-\frac{bQ^2\hat{G}^3(Q^2)}{2(Q^2+m^2)^2}
\end{equation}
where a prime indicates differentiation with respect to $Q^2$.
If we now define $t$ as:
\begin{equation}
\label{redeft}
t=\ln (\frac{Q^2+m^2}{\Lambda^2}),
\end{equation}the differential equation can be written as an extension of
the massless equation:
\begin{equation}
\label{massdiffeq}
\hat{G}_{tt}+\hat{G}_t=-\frac{b}{2}\hat{G}^3(1-\frac{m^2e^{-t}}{\Lambda^2}).
\end{equation}
The difference from the massless equation is that at $Q=0$ the variable $t$ 
is finite, not $-\infty$.

It is not difficult to check from the integral equation (\ref{approxint})
that  at large momenta only the $J_2$ term is leading, yielding   the exact
leading behavior $\hat{G}(Q^2)\rightarrow 1/(b\ln Q^2)^{1/2}$ at large
$Q^2$.   The $J_1$ term in Eq.~(\ref{approxint}) is $\mathcal{O}[(\ln
Q^2)^{-3/2})]$ and non-leading.  In neither the original toy model nor at
present are we interested in such non-leading terms, so we define our new
toy  model by dropping the $J_1$ term.  Note also that
$J_1=\mathcal{O}(Q^4)$ at small $Q^2$, while $J_2=\mathcal{O}(1)$, since we
require that $\hat{G}(0)\neq 0$; the missing $J_1$ term is non-leading in
the infrared as well.  [One can show that the
$\mathcal{O}(Q^4)$ terms in $J_1$ and $J_2$ cancel in the full equation,
leaving corrections to $\hat{G}(0)$ of $\mathcal{O}(Q^6)$.  In fact,   the
exact solution $\hat{G}$  of Eq.~(\ref{approxint}) with both terms show the
self-consistent behavior leading correction at small momentum:
\begin{equation}
\label{smallq}
\hat{G}(Q^2)\simeq \hat{G}(0)-\frac{\hat{G}(0)^3}{12}(\frac{Q^2}{m^2})^3
+\dots ]
\end{equation}

  Evidently the resulting
integral equation with no $J_1$ term satisfies a first-order differential
equation, which is
just Eq.~(\ref{massdiffeq}) without the $\hat{G}_{tt}$ term.  This equation
  has   the exact   solution $\hat{G}_1$:
\begin{equation}
\label{masseq}
 \frac{1}{\hat{G}^2_1(Q^2)}=b[\ln
(\frac{Q^2+m^2}{\Lambda^2})+\frac{m^2}{Q^2+m^2}].
\end{equation}
One can check
that the beta-function coming from $\hat{G}_1$ has  not only the usual
$-b\hat{G}_1^3$ term but also terms involving non-perturbative quantities
such as $\exp [-1/(b\hat{G}_1^2)]$.  This approximation already shows a
mass $m_0$ at which $\alpha_s(0)$ is singular:   $m_0/\Lambda =
e^{-1/2}\approx 0.61$; smaller values
lead to a pole in $\hat{G}_1$.     (We have done a quick numerical study of
the
differential equation Eq.~(\ref{massdiffeq}) of the modified model, and
found $m_0/\Lambda \approx 0.66$.)   The actual critical mass $m_c$ might be
about twice as large as $m_0$, based on our experience with the propagator
model.

\subsection{\label{toynomass} Is a gluon mass really necessary?}

\begin{figure}
\includegraphics[width=5in]{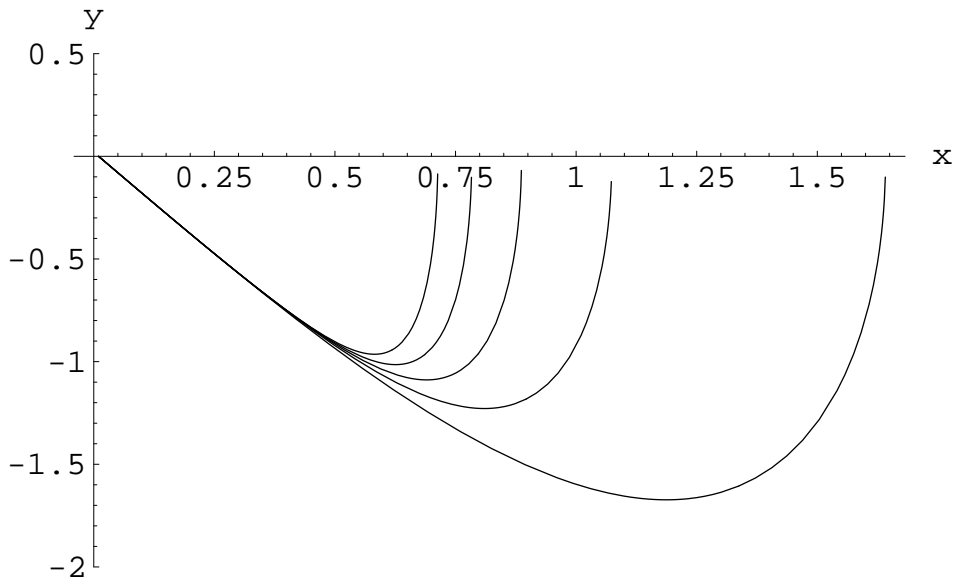}
\caption{\label{fig2} Solutions to the differential equation (\ref{kdiffeq})
for $K$=1 (right-most curve) to $K$=5 (left-most curve).  Here
$x=\sqrt{b}g$ and $y=\sqrt{b}\beta$.  }
\end{figure}

One may ask whether higher-order effects in a massless theory can somehow do
away with the need for a gluon mass.  To study this possibility,  let us
now extend the basic one-loop   beta-function equation
(\ref{nomassbeta}) to mimic multi-loop vertex contributions.
   Of course, we can only aspire to qualitative accuracy, looking for
mechanisms of limits on $g^2$ rather than for accurate values of these
limits.  These models are essentially those of \cite{corn112} devised for
$\phi^3_6$, with coupling $G$.  In this theory it has been shown that the
sum of all $K$-loop vertex graphs is always positive and grows like
$K!(cG^{2})^K$, where $c$ is a positive number.   We conjecture that the
same holds for $d=4$ NAGTs, and that an appropriate extension of the
beta-function differential equation to higher loops is:
\begin{equation}
\label{kdiffeq}
 y(1+\frac{1}{2}y')=-x\sum_{J=1}^KJ!x^{2J}
\end{equation}
where
\begin{equation}
\label{defxy}
x=\sqrt{b}g,\;\;y(x)=\sqrt{b}\beta (g)
\end{equation}
 and the prime denotes an $x$ derivative.   One can verify that the
beta-function solving this equation grows factorially like $K!(cg^2)^K$ for
some $c$, whose value is irrelevant to our investigation.   Our only
interest is in seeing what happens when the right-hand side of the vertex
Schwinger-Dyson equation has terms that grow factorially with loop number
$K$.  In fact, we have tested the sensitivity of Eq.~(\ref{kdiffeq}) to
fairly major changes in the coefficients on the right-hand side, and find
little change from the solutions to Eq.~(\ref{kdiffeq}) as shown in
Fig.~\ref{fig2}.  

\begin{table}
\caption{\label{table1} {\bf Upper-bound couplings for various $K$}}
\begin{center}
\begin{tabular}{|r||l|l|l|l|l|}\hline
$K$ = &  1 & 2 & 3 & 4 & 5\\ \hline  
$\alpha_c$  = & 3.1 & 1.3 & 0.90 & 0.70 & 0.58   \\ \hline
\end{tabular}
\end{center}
\end{table}

Table 1 show the upper bounds $\alpha_c$ for quarkless QCD, defined as the
value of
$g_c^2/(4\pi )$ at which the beta-function crosses the real axis, for
various values of $K$.  One should expect the $\alpha_c$ values to decrease
as $K$ increases, since otherwise the right-hand side of the differential
equation will grow too large to balance the left-hand side.  For all values
of $K$ there is a singularity  of the form $\beta \sim (g_c-g)^{1/2}$,
where $\beta$
has infinite slope at $g=g_c$ and then turns imaginary for $g>g_c$.   
Note that there seems to be numerical convergence toward a value near 0.5,
not far from ``best" estimates based on the one-dressed-loop pinch
technique with a mass. But there is always a singularity.  

Perhaps some way of summing the non-Borel-summable series of
Eq.~(\ref{kdiffeq}) would remove this singularity.  We have tried, again in
the spirit of \cite{corn112}, ``regulating" the all-orders behavior with a
principal part integral form of the  Borel  integral corresponding to the
sum in Eq.~(\ref{kdiffeq}), using:
\begin{equation}
\label{nomassbeta2}
y(1+\frac{1}{2}y')=L(x).
\end{equation}
with:
\begin{equation}
\label{allorderdiffeq}
L(x) = -\frac{x^2}{\pi^{1/2}}\int_0^{\infty}
      \frac{d\alpha}{\alpha^{1/2}}\{\frac{e^{-\alpha /x^2} - e^{-1/x^2} }
          {1 - \alpha}\}.
\end{equation}
The equation for the vertex itself, analogous to Eq.~(\ref{massless}), is:
\begin{equation}
\label{massless1}
\hat{G}_{tt}+\hat{G}_t=b^{-1/2}L(b^{1/2}\hat{G}).
\end{equation}
Actually, some form of principal-part regulation is demanded by dynamical
boson mass generation \cite{corn112}, but we  need not inquire further into
that here.  We can understand the basic behavior of the equation by looking
at the degree to which factorial growth at large coupling is tamed by the
principal-part prescription.
The power-series expansion of $L(x)$ is
\begin{equation}
\label{xseries}
L(x)=-\frac{1}{\sqrt{\pi}}\sum_{K=1}x^{2K+1}\Gamma
(K-\frac{1}{2})+\mathcal{O}
(e^{-1/x^2})
\end{equation}
and its asymptotic behavior at infinity is $-2x$.   
This is very different from the finite-$K$ models of Eq.~(\ref{kdiffeq}). 
In these, the increasingly-strong growth with $x$ of the right-hand side as
$K$ gets larger means that any singularity occur at smaller values of $x$. 
But for equations with the $L(x)$ source there is no such movement toward
smaller couplings because $L(x)$ is not growing rapidly at large $x$.  In
the beta-function equation
(\ref{nomassbeta2}) one easily finds the large-$x$ asymptotic behavior
$y(x)\rightarrow -cx  $ with $c=-1+\sqrt{5}$.  So instead of generating
a finite-$g$ singularity, the beta-function turns from $-bg^3$ behavior
near the origin to linear  at large enough $g$; numerical simulations
confirm
this.  For the massless vertex equation (\ref{massless1}) there is still a
singularity in the infrared, so that $\hat{G}$ gets large and the large $x$
behavior of $L(x)$ matters.  In this regime the massless equation
(\ref{massless1})
becomes linear, and one finds unphysical behavior of the type $(\ln
Q^2)^{-1/2} 
cos (\sqrt{3}\ln Q^2)$.  So massless vertex dynamics is not regularized by
the
specific behavior of $L(x)$.  Mass is important not only for the right-hand
sides of the vertex and beta-function equations, as summarized by the
function $L$, but it is important in the left-hand side of such equations,
as we show below.  In fact, the mass damping is so strong that it is
probably unnecessary to worry about vertex graphs of very high order;  
$\hat{G}$ does not get large enough to probe the asymptotic limit of
$L(b\hat{G}^2)$, as it does for the massless case. 

\section{Summary and conclusions}

We argue that although the gauge-invariant PT propagator does not obey the
K-L representation, it is the product of two factors that do
have the required positive spectral functions.  We show that this holds
true for an analytic approximation to the one-dressed-loop PT  propagator
equation, and that this approximation implies a critical mass $m_c$ such
that the true dynamical mass $m$ must exceed $m_c$ or spurious
singularities arise.  We construct an analytically-soluble toy model of
one-dressed-loop PT three-gluon vertex model with a cubic non-linearity and
show how the positivity of the factor $\hat{H}$ in the PT propagator plays
an
essential role in the vertex dynamics, in particular the  occurrence of a
mass value which must be exceeded by $m$ to avoid unwanted singularities. 
We argue that higher-order, even regulated all-order, extensions of the
{\em massless} toy model equations do not remove these singularities.


\begin{thebibliography}{99}
\bibitem{corn82}  J.~M.~Cornwall, Phys.\ Rev.\  D {\bf 26}, 1453 (1982).
\bibitem{mandog}  J.~E.~Mandula and M.~Ogilvie, Phys.\ Lett.\  B {\bf 185},
127 (1987).
\bibitem{auog} C.~A.~Aubin and M.~C.~Ogilvie, Phys.\ Lett.\  B {\bf 570}, 59
(2003).
\bibitem{bowetal} P.~O.~Bowman {\it et al.}, Phys.\ Rev.\  D {\bf 76},
094505 (2007). 
\bibitem{amn}   A.~C.~Aguilar, A.~Mihara and A.~A.~Natale, Phys.\ Rev.\  D
{\bf 65}, 054011 (2002).
\bibitem{cornhou}   J.~M.~Cornwall and W.~S.~Hou, Phys.\ Rev.\  D {\bf 34},
585 (1986).
\bibitem{papag}   A.~C.~Aguilar, D.~Binosi and J.~Papavassiliou, PoS {\bf
LC2008}, 050 (2008).
\bibitem{abp}  A.~C.~Aguilar, D.~Binosi and J.~Papavassiliou,
 Phys.\ Rev.\  D {\bf 78}, 025010 (2008).
\bibitem{corn136}  J.~M.~Cornwall, Phys.\ Rev.\  D {\bf 76}, 025012 (2007).
\bibitem{shirsol}  D.~V.~Shirkov and I.~L.~Solovtsov, Phys.\ Rev.\ Letters
{\bf 79}, 1209 (1997).
\bibitem{shirk}  D.~V.~Shirkov,  Phys.\ Atom.\ Nucl.\  {\bf 62}, 1928 (1999)
  [Yad.\ Fiz.\  {\bf 62}, 2082 (1999)].
\bibitem{papnest}    A.~V.~Nesterenko and J.~Papavassiliou,
  Int.\ J.\ Mod.\ Phys.\  A {\bf 20}, 4622 (2005);  Phys.\ Rev.\  D {\bf
71}, 016009 (2005).
\bibitem{corn99} J.~M.~Cornwall and J.~Papavassiliou, Phys.\ Rev.\  D {\bf
40}, 3474 (1989).
\bibitem{papbin} D.~Binosi and J.~Papavassiliou, Phys.\ Rev.\  D {\bf 66},
111901 (2002); J.~Phys.~G {\bf 30}, 203 (2004).
\bibitem{abbott} L.~F.~Abbott, Nucl.\ Phys.\  B {\bf 185}, 189 (1981).
\bibitem{corn77}  J.~M.~Cornwall, in ``Deeper Pathways in High-Energy
Physics", ed. B.~Kursonoglu, A.~Perlmutter, and L.~Scott, p. 683 (Plenum,
New York, 1977).
\bibitem{bdf}  V.~L.~Baltar, H.~G.~Dosch and E.~Ferreira, arXiv:0907.5310
[hep-ph].
\bibitem{papbin2} D.~Binosi and J.~Papavassiliou, Phys. Rev. D {\bf 77},
061702 (2008);  
JHEP {\bf 0811}, 063 (2008).
\bibitem{lavelle} M.~Lavelle, Phys.\ Rev.\  D {\bf 44}, R26 (1991).
\bibitem{alex1} C.~Alexandrou, P.~de Forcrand and E.~Follana,
  Phys.\ Rev.\  D {\bf 63}, 094504 (2001).
\bibitem{alex2}C.~Alexandrou, P.~de Forcrand and E.~Follana,
  Phys.\ Rev.\  D {\bf 65}, 114508 (2002).
\bibitem{bouc1} Ph.~Boucaud {\it et al.},
  arXiv:hep-ph/0507104.
\bibitem{bowm1}  P.~O.~Bowman, U.~M.~Heller, D.~B.~Leinweber,
M.~B.~Parappilly and A.~G.~Williams,
  Phys.\ Rev.\  D {\bf 70}, 034509 (2004).
\bibitem{bouc2}  Ph.~Boucaud {\it et al.},
  arXiv:hep-lat/0602006.
\bibitem{silva1}  A.~Cucchieri, T.~Mendes, O.~Oliveira and P.~J.~Silva,
  Phys.\ Rev.\  D {\bf 76}, 114507 (2007).
\bibitem{silva2} O.~Oliveira and P.~J.~Silva,
  Eur.\ Phys.\ J.\  A {\bf 31}, 790 (2007).
 \bibitem{bogo1} I.~L.~Bogolubsky, E.~M.~Ilgenfritz, M.~M\"uller-Preussker
and
A.~Sternbeck,
  PoS {\bf LAT2007}, 290 (2007).
\bibitem{stern} A.~Sternbeck, L.~von Smekal, D.~B.~Leinweber and
A.~G.~Williams,
  PoS {\bf LAT2007}, 340 (2007).
\bibitem{bogo2} I.~L.~Bogolubsky, V.~G.~Bornyakov, G.~Burgio,
E.~M.~Ilgenfritz, M.~M\"uller-Preussker, P.~Schemel and V.~K.~Mitrjushkin,
  PoS {\bf LAT2007}, 318 (2007).
\bibitem{oliv} O.~Oliveira, P.~J.~Silva, E.~M.~Ilgenfritz and A.~Sternbeck,
  PoS {\bf LAT2007}, 323 (2007).
\bibitem{bowm2} P.~O.~Bowman {\it et al.},
  Phys.\ Rev.\  D {\bf 76}, 094505 (2007).
\bibitem{ilge2} E.~M.~Ilgenfritz, M.~M\"uller-Preussker, A.~Sternbeck,
A.~Schiller and I.~L.~Bogolubsky,
  Braz.\ J.\ Phys.\  {\bf 37}, 193 (2007).
\bibitem{cucch1} A.~Cucchieri and T.~Mendes,
  PoS {\bf LAT2007}, 297 (2007).
\bibitem{cucch2} A.~Cucchieri and T.~Mendes,
  Phys.\ Rev.\ Lett.\  {\bf 100}, 241601 (2008).
\bibitem{bouc3} Ph.~Boucaud, J.~P.~Leroy, A.~Le Yaouanc, J.~Micheli, O.~Pene
and J.~Rodriguez-Quintero,
  JHEP {\bf 0806}, 099 (2008).
\bibitem{zhang}  Zhang, Y-B, J-L Ping, X-F Lu, and H-S Zong, 
Comm. Theor. Phys. (Beijing) {\bf 50}, 125 (2008).
\bibitem{gong} M.~Gong, Y.~Chen, G.~Meng and C.~Liu, arXiv:0811.4635
[hep-lat], December, 2008.
\bibitem{bims}  I.~L.~Bogolubsky, E.~M.~Ilgenfritz, M.~M\"uller-Preussker
and
A.~Sternbeck, arXiv:0901.0736 [hep-lat].
\bibitem{dfh}  S.~D.~Drell, A.~C.~Finn and A.~C.~Hearn, Phys.\ Rev.\  {\bf
136}, B1439 (1964).
   \bibitem{corn112}  J.~M.~Cornwall and D.~A.~Morris, Phys.\ Rev.\  D {\bf
52}, 6074 (1995).
\end{thebibliography}
\end{document}